  \documentclass[letter]{IEEEtran}
\usepackage{algorithm,algpseudocode,graphicx}  
\usepackage{amsfonts}
\usepackage{tabularx}
\usepackage{subfigure}
\usepackage{amssymb}
\usepackage{booktabs} 
\usepackage{multirow} 
\usepackage{epsfig}
\usepackage{epstopdf} 
\providecommand{\e}[1]{\ensuremath{\times 10^{#1}}}
\usepackage[noadjust]{cite}

\newtheorem{remark}{Remark}

\newtheorem{lemma}{Lemma}

\usepackage[cmex10]{amsmath}
\allowdisplaybreaks  
   \usepackage{color}
\begin{document} 
\title{ Optimal Green Hybrid Attacks in Secure IoT}  
 \author 
  {Bhawna Ahuja,~\IEEEmembership{Student~Member,~IEEE,}  Deepak Mishra,~\IEEEmembership{Member,~IEEE,}  and Ranjan Bose,~\IEEEmembership{Senior~Member,~IEEE} 
  \thanks{B. Ahuja  and R. Bose are  with the Bharti School of  Telecommunication Technology and Management, IIT Delhi,110016 New Delhi.  R. Bose is also affiliated with IIIT Delhi.   e-mail: (bhawna.ahuja,rbose)@iitd.ac.in}
  \thanks{D. Mishra is with the Department of
  School of Electrical Engineering and Telecommunications, UNSW Sydney, NSW 2052, Australia
   (e-mail: d.mishra@unsw.edu.au).}
}
 \maketitle

 \begin{abstract}
  In pursuance to understand the behavior of a potential green hybrid attacker in secure internet-of-things (IoT), this letter investigates optimal energy utilization from the attacker's viewpoint. Specifically, we propose a novel framework for optimizing the efficacy of a hybrid attacker, possessing capability to both eavesdrop and jam, while considering the underlying energy consumption.  In particular, we maximize the attacker energy efficiency (AEE) in  secure IoT by deriving the analytical solutions for jointly global-optimal attacking mode, eavesdropping rate, and jamming power. Numerical results, validating  analytical claims, reveal that the proposed green design can offer about $45 \%$ improvement in average AEE over the relevant benchmarks. 
 \end{abstract}  \vspace{-1mm}
 \vspace{-2mm}
 
\section{Introduction}\label{sec:introduction} 
Physical layer security (PLS) is a promising approach for  investigating the security threats in resource-constrained setups like internet of things (IoT) \cite{survey18}, especially with 
an intelligent hybrid attacker capable of carrying both eavesdropping and jamming ~\cite{TWC19Hybrid}. 
Along with the security, energy efficiency (EE)   has also become important
from both user and attacker's perspective \cite{jamming2019}.  
 %
{This has led to a timely demand for investigating 
a green hybrid attacker
which may be  a potential threat for  practical IoT applications
 like smart home and smart city.}

Recent studies have shown interest on optimizing the EE of jamming attacker
\cite{Learjam,jamming2019}. 
 Whereas optimal attacking strategies for hybrid attacker are designed in \cite{HA18,Garnaev} to achieve  maximum degradation in the secrecy rate, however EE aspect has been neglected.
 Further,  decoding-based energy-expenditures in eavesdropping and  practical circuit-level consumption during jamming have been ignored while considering  energy consumption at attacker 
 \cite{Learjam,jamming2019,HA18,Garnaev,EEE19}.
{
In contrast, for the IoT applications using low power wireless communication technologies e.g. Bluetooth Low Energy (BLE), ZigBee, LoRa, SigFox etc.,
active power at receiver becomes non-negligible in comparison to that at  transmitter   \cite{computing}. This can have a significant effect on power consumption strategies, hence  deciding whether to eavesdrop or jam, becomes a nontrivial task for the attacker.}

 {We address this challenge by proposing novel energy-aware attacking strategies for hybrid attackers using theoretical analyses of PLS  in secure IoT.} 
Major contributions include: (i) First-time characterization of the  secrecy rate degradation  arising out of  underlying joint energy-aware eavesdropping and  jamming. 
The key aspect is to consider energy consumption during both attacker modes and propose a new metric called attacker energy efficiency (AEE) that can fairly model the combined degradation effect. 
(ii) Derivation of a semi-closed form expression for  joint global-optimal attacking mode, eavesdropping rate  and jamming power in jamming mode yielding  maximal AEE. 
(iii) Practically-motivated numerical investigation for validating the proposed analysis, that includes tight analytical approximation for global-optimal jamming power, and quantifying the achievable AEE gains over relevant benchmarks. 
\color{black}

\section{ Novel Green Hybrid Attacker Model }\label{sec:model} 
\subsection{Network  Topology and Channel Model} \label{sec:topology}
We consider secure data transmission in an IoT system,  wherein an information source $\mathcal{S}$ communicates with a legitimate user $\mathcal{U}$ using power $P_\mathcal{S}$  in the presence  of an attacker $\mathcal{A}$. { Here, $\mathcal{A}$ is a malicious, but smart IoT node possessing the dual capability of eavesdropping and jamming, though having a limited energy budget $P_m$.} $\mathcal{A}$ works in a half-duplex mode \cite{TWC19Hybrid} and  while eavesdropping, it turns on its receiver, and attempts to tap the legitimate  transmissions of $\mathcal{S}$. Alternatively, when $\mathcal{A}$  acts as a jammer, it sends a jamming signal at power $P_{\mathcal{J}}$ to impair the legitimate receptions at  $\mathcal{U}$.  The corresponding channel gains over   $\mathcal{S}$-$\mathcal{U}$,    $\mathcal{S}$-$\mathcal{A}$  and   $\mathcal{A}$-$\mathcal{U}$ links are respectively denoted by $\mathrm{g}_{_{\mathcal{S}\mathcal{U}}}$,  $\mathrm{g}_{_{\mathcal{S}\mathcal{A}}}$ and $\mathrm{g}_{_{\mathcal{A}\mathcal{U}}}$. These links are assumed to be  perfectly known at  $\mathcal{A}$ \cite{EEE19} for each coherence  block of unit duration.


  
  \subsection{  Degraded Secrecy Rate  } \label{sec:DSR}
  The respective signal-to-noise ratios (SNRs) 
  for  %
  links $\mathcal{S}$-$\mathcal{U}$  and  $\mathcal{S}$-$\mathcal{A}$  are given by $\gamma_{_{\mathcal{S}\mathcal{U}}}=\frac{P_{\mathcal{S}}\mathrm{g}_{_{\mathcal{S}\mathcal{U}}}}{\sigma^{2}}$ and $\gamma_{_{\mathcal{S}\mathcal{A}}}=\frac{P_{\mathcal{S}}\mathrm{g}_{_{\mathcal{S}\mathcal{A}}}}{\sigma^{2}}$ where  $\sigma^2$ is the received noise power  at $\mathcal{U}$ and $\mathcal{A}$. Accordingly, respective  communication rates  or spectral efficiencies   $R_\mathcal{U}$ and  $R_\mathcal{A }$ over  links $\mathcal{S}$-$\mathcal{U}$  and  $\mathcal{S}$-$\mathcal{A}$ in (bps/Hz) are calculated as  $ R_\mathcal{U}=\log_{2}(1+ \gamma_{_{\mathcal{S}\mathcal{U}}})$ and $R_\mathcal{A }=\log_{2}(1+ \gamma_{_{\mathcal{S}\mathcal{A}}})$.
  Now, the secrecy rate   $R_{\mathcal{E}}$ for $\mathcal{S}$-$\mathcal{U}$ communication under eavesdropping attack by $\mathcal{A}$ \cite {survey18} is : 
\begin{align}\label{eq: C_E}R_{\mathcal{E}}  & =\max(0,R_{\mathcal{U}}- R_{\mathcal{A}}).
\end{align}
Similarly under $\mathcal{A}$'s jamming mode,  secrecy rate $R_{\mathcal{J}}$  
\cite{ TWC19Hybrid} is:
    \begin{align}\label{eq: C_J} 
    \hspace{-3mm}
    R_{\mathcal{J}} & =\log_{2}\left(1+ \frac{P_{\mathcal{S}}  \mathrm{g}_{_{\mathcal{S}\mathcal{U}}}}{P_{\mathcal{J}} \mathrm{g}_{_{\mathcal{A}\mathcal{U}}} +\sigma^{2}}\right) =\log_{2}\left(1+ \frac{\gamma_{_{\mathcal{S}\mathcal{U}}}}{1+ \gamma_{_{\mathcal{A}\mathcal{U}}}}\right), \hspace{-2mm}\end{align} 
    where  $\gamma_{_{\mathcal{A}\mathcal{U}}}=\frac{P_{\mathcal{J}}\mathrm{g}_{_{\mathcal{A}\mathcal{U}}}}{\sigma^{2}}$ represents the SNR of $\mathcal{A}-\mathcal{U}$ link.
   
    Now, as  $\mathcal{A}$ intends to   maximize deterioration to the  secrecy  rate of $\mathcal{U}$,  we express  this deterioration to  $\mathcal{U}$'s secure performance  via  degraded secrecy  rate.
This metric is quantified  as difference of    achievable rate during $\mathcal{S}$-$\mathcal{U}$ communication in absence of $\mathcal{A}$ and the corresponding  secrecy rate $R_{\mathcal{E}}$ or $ R_{\mathcal{J}}$   in presence of $\mathcal{A}$.  Hence, respective  degraded secrecy  rate $R_{\text{{D}}{\mathcal{E}}}$  and $R_{\text{{D}}{\mathcal{J}}}$ for eavesdropping and jamming  are defined  as:
 \begin{align}\label{eq: C_LE}
  R_{\text{{D}}{\mathcal{E}}}  & =  {R_\mathcal{U}-R_{\mathcal{E}}} 
                           = \min( R_\mathcal{A},R_\mathcal{U}),
 \end{align}%
  \begin{align}\label{eq: C_LJ}
 R_{\text{D}\mathcal{J}}  & =  {R_\mathcal{U}-R_{\mathcal{J}}} 
         =\log_{2}\left(\frac{(1+ \gamma_{_{\mathcal{S}\mathcal{U}}})(1+ \gamma_{_{\mathcal{A}\mathcal{U}}})}{1+\gamma_{_{\mathcal{S}\mathcal{U}}}+\gamma_{_{\mathcal{A}\mathcal{U}}}}\right).
         \end{align}
  \subsection{ Consumption Model for Eavesdropping and Jamming } \label{sec:consumption}
 {As in eavesdropping, $\mathcal{A}$ has to  receive and decode the signal, we have followed the commonly used model  for computing power consumption at receiver \cite {Dilemma_Deepak,Access, energyModel2019},  and references therein. For a unit block duration of 1 s, the energy or power  consumption at $\mathcal{A}$  in receiving mode   is given by:
      $  {P_\text{fr}+ \rho_{_\text{d}}\hspace{.5mm} r_\mathcal{A} }  $.
  Here, $ P_\text{fr}$ is the static  consumption required for  receiving at $\mathcal{A}$,  $\rho_{_\text{d}}$ represents the dynamic unit rate decoding power consumption  and $r_\mathcal{A}$ represents  the  decoding rate of $\mathcal{A}$.}
  It may be noted here  
  the decoding rate $r_\mathcal{A}$ chosen at $\mathcal{A}$ has to satisfy the constraint $r_\mathcal{A}\leq R_{\mathcal{A}}$, because $R_{\mathcal{A}}$ is the maximum  rate over  $\mathcal{S}$-$\mathcal{A}$ link as available for eavesdropping.
   { Similarly, as a jammer, $\mathcal{A}$ transmits a jamming signal at power $P_{\mathcal{J}}$, adopting conventional power consumption model for transmitter \cite{Dilemma_Deepak,SEE18}}. Therefore, the power consumption in jamming mode is modelled as: $ P_\text{ft}+(P_{\mathcal{J}}/\nu)$.
  Here, $P_\text{ft}$ is static power consumption at transmitting circuitry of $\mathcal{A}$   and $\nu $ is  efficiency of power amplifier at $\mathcal{A}$. 
 \subsection{Objective Formulation} \label{Obj}
We  introduce a new performance metric called Attacker Energy Efficiency (AEE) defined as a ratio of the degraded secrecy  rate to the total power consumption at $\mathcal{A}$ in achieving it.
 { It is worth to mention that  being the decided decoding rate at $\mathcal{A}$, $r_\mathcal{A}$ reflects the effective eavesdropping,  henceforth may also be known as  eavesdropping rate. Noting that degradation in secrecy rate caused by eavesdropping  is upper bounded by $R_\mathcal{U}$,  $r_\mathcal{A}$ actually characterise the effective degraded secrecy rate under eavesdropping mode which is eventually defined as $ \min( r_\mathcal{A}, R_\mathcal{U})$. } It is upper bounded by $R_{\text{{D}}{\mathcal{E}}} $ given by \eqref{eq: C_LE} due to  constraint $r_\mathcal{A}\leq R_{\mathcal{A}}$. 
%
 Now AEE $\eta_{\mathcal{E}}\{  r_\mathcal{A}\}$ for eavesdropping,  and $\eta_{\mathcal{J}}\{P_{\mathcal{J}}\}$ for jamming using  \eqref{eq: C_LJ} are respectively given as: 
 \begin{equation}\label{AEE_e}          
 \eta_{\mathcal{E}}\{  r_\mathcal{A}\}= \dfrac{\min(r_\mathcal{A}, R_\mathcal{U})}{P_\text{fr}+ \rho_{_\text{d}}\hspace{.5mm} r_\mathcal{A}}, \quad \eta_{\mathcal{J}}\{P_{\mathcal{J}}\}= \dfrac{R_{\text{D}\mathcal{J}}}{P_\text{ft}+(P_{\mathcal{J}}/\nu)}.
  \end{equation}
  
  {For incorporating overall degradation in secrecy rate due to both
 eavesdropping and jamming modes of $\mathcal{A}$, AEE $\eta\{ \alpha, r_\mathcal{A}, P_\mathcal{J} \}$ can be defined as a ratio of weighted sum of degraded secrecy rate for underlying attacking modes to the weighted sum of their respective power consumption.}
\begin{equation} \label{AEE}
\hspace{-11mm}\eta\{ \alpha, r_\mathcal{A}, P_\mathcal{J} \}= {\dfrac{\alpha \hspace{0.2mm}{\min(r_\mathcal{A}, R_\mathcal{U})+ (1-\alpha) R_{\text{D}\mathcal{J}} }}{\alpha (P_\text{fr}+ \rho_{_\text{d}}\hspace{.2mm} r_\mathcal{A})+ (1-\alpha) (P_\text{ft}+(P_{\mathcal{J}}/\nu))}}, \hspace{-6mm}
\end{equation}
where $\alpha$ is the fraction of unit block duration assigned for eavesdropping and  $(1-\alpha)$ is  allocated for jamming.

 We aim to  maximise AEE $\eta\{ \alpha, r_\mathcal{A}, P_\mathcal{J} \}$ by jointly optimising $\alpha, r_\mathcal{A}, P_\mathcal{J}$ as in optimization problem  (P0),
subject to eavesdropping rate constraints ($C1$-$C2$),  jamming  power constraints ($C3$-$C4$), normalization constraints on $\alpha$ ($C5$-$C6$), and total power constraint ($C7$) at $\mathcal{A}$,  is:
\begin{align} \label{P0}
&\text{(P0)}:\;\underset{r_\mathcal{A}, P_{\mathcal{J}}, \alpha}{\text{max}} \eta\{ \alpha, r_\mathcal{A}, P_\mathcal{J} \},\quad\text{subject to:}\nonumber \\
 & C1: {r_\mathcal{A}}\le R_{\mathcal{A}},\;C2: {r_\mathcal{A}}\ge 0,\; C3:P_{\mathcal{J}}\le P_{\mathcal{J}m},\nonumber\;\\
&C4: P_{\mathcal{J}}\ge 0,\;
\hspace{3mm} 
C5:\alpha \ge 0,\;   
 \hspace{2mm} 
C6: \alpha \le 1,\nonumber\;\\
 &C7: {\alpha (P_\text{fr}+ \rho_{_\text{d}}\hspace{.5mm} r_\mathcal{A})+ (1-\alpha) (P_\text{ft}+(P_{\mathcal{J}}/\nu))} \le P_{m}. \;  \hspace{-3mm}\nonumber
\end{align} 
Here, $P_{\mathcal{J}m}$ in $C3$ is maximum allowed jamming power.
Next, we present jointly global-optimal solution for ({P0}).
\section{Proposed  Solution Methodology } \label{sec:SM}
  The original problem ({P0}) is a nonlinear non-convex  optimization problem due to non-concavity of objective function $\eta\{ \alpha, r_\mathcal{A}, P_\mathcal{J} \}$ in $\alpha $, $ r_\mathcal{A}$ and $P_{\mathcal{J}}$.
 {So, we first propose an  equivalent transformation strategy involving  decomposition of (P0) into three sub-problems
and then obtain the   jointly global-optimal solution by using outcomes of these $3$ sub-problems.}
\subsection{Optimal Mode Selection } \label{sec:P1}
The sub-problem (P1) to obtain optimal $\alpha $ that maximizes $\eta\{ \alpha\}$ for a given $ r_\mathcal{A}$ and $P_{\mathcal{J}}$, is defined as:
\begin{eqnarray}\label{P1}
 &(\text{P1}):\;\underset{\alpha}{\text{max}} \;\eta\{ \alpha\},
 &\text{subject to:}\; C5, C6, C7. \nonumber
 \end{eqnarray}
 The solution of ({P1}) is presented by Lemma 1  as given below.
  \begin{lemma}\label{lem:alpha}  The global-optimal value of $\alpha$ is either  1 or 0. 
 \end{lemma}
 \begin{IEEEproof}
 Depending on input parameters,   $\eta\{\alpha  \}$ will  either be increasing or decreasing function of $\alpha$ as shown next by  taking first order derivative of $\eta\{\alpha  \}$ with respect to $\alpha$:
  \begin{eqnarray}\label{FDP1}
   \frac{\partial \eta\{\alpha  \}}{\partial \alpha}\hspace{-1mm} &= \hspace{-1mm}\frac {(P_\text{fr}+ \rho_{_\text{d}}\hspace{.2mm} r_\mathcal{A}) (P_\text{ft}+(P_{\mathcal{J}}/\nu)) ({\frac{\min(r_\mathcal{A}, R_\mathcal{U})}{P_\text{fr}+ \rho_{_\text{d}}\hspace{.2mm} r_\mathcal{A} }- \frac{ R_{\text{D}\mathcal{J}}}{P_\text{ft}+(P_{\mathcal{J}}/\nu)} })}{{( \alpha (P_\text{fr}+ \rho_{_\text{d}}\hspace{.2mm} r_\mathcal{A})+ (1-\alpha)(P_\text{ft}+(P_{\mathcal{J}}/\nu ))}^2} \nonumber\\
   & \stackrel{a}{=}\frac {(P_\text{fr}+ \rho_{_\text{d}}\hspace{.2mm} r_\mathcal{A}) (P_\text{ft}+(P_{\mathcal{J}}/\nu)) ({\eta_\mathcal{E}\{r_\mathcal{A}\}- \eta_\mathcal{J}\{P_\mathcal{J}\} })}{{( \alpha P_\text{fr}+ \rho_{_\text{d}}\hspace{.2mm} r_\mathcal{A} + (1-\alpha)P_\text{ft}+(P_{\mathcal{J}}/\nu) )}^2},
 \end{eqnarray}
  where $a$ is obtained using (5). 
Being $\frac{\partial \eta\{\alpha  \}}{\partial \alpha}>0$ and $\frac{\partial \eta\{\alpha  \}}{\partial \alpha}\leq 0$  respectively 
for $ \eta_\mathcal{E}\{r_\mathcal{A}\}> \eta_\mathcal{J}\{P_\mathcal{J}\}$ and $ \eta_\mathcal{E}\{r_\mathcal{A}\}\leq \eta_\mathcal{J}\{P_\mathcal{J}\}$, it is observed that $\eta\{ \alpha\}$ is an increasing
 function of $\alpha$ in the former case and decreasing function otherwise. 
Since  $\eta_{\mathcal{E}}\{  r_\mathcal{A}\}$ and $\eta_{\mathcal{J}}\{P_{\mathcal{J}}\}$  do not vary with $\alpha$, both can be easily compared to find  optimal solution  existing at extreme points of  $ { \alpha}$, i.e.,   1  or 0.
In other words, assigning only one attacking mode  either eavesdropping or jamming  is optimal as follows:
\begin{equation}\label{optalpha}\alpha^{\star} = \begin{cases} 1 \quad \text{(Eavesdropping)},& \eta_\mathcal{E}\{r_\mathcal{A}\}> \eta_\mathcal{J}\{P_\mathcal{J}\}\\
0 \quad \text{(Jamming)}, & \text{otherwise},
\hspace{-7mm}
\end{cases} 
\end{equation} \hspace{-3mm} where $\alpha^{\star}$ is optimal solution of (P1).  \end{IEEEproof}

{\begin{remark}
Noting the two possible values of  $\alpha^\star$,  objective function $\eta\{ \alpha, r_\mathcal{A}, P_\mathcal{J} \}$ of  problem (P0)  equivalently reduces to $\eta_{\mathcal{E}}\{  r_\mathcal{A}\}$ for $\alpha^\star=1$ or $\eta_{\mathcal{J}}\{P_{\mathcal{J}}\}$ for $\alpha^\star=0$. It shows that $r_\mathcal{A}$ and $P_\mathcal{J}$ are decoupled from each other at optimal value of $\alpha$ that enables each variable to be individually optimised. 
 \end{remark}
}
 \subsection{Optimal Eavesdropping  Rate}\label{sec:P4}
{For $\alpha=1$, (P0) is reduced to  (P2) where  objective function $\eta_{\mathcal{E}}\{  r_\mathcal{A}\}$ represents AEE for eavesdropping only. }
\begin{eqnarray}\label{P2}
 (\text{P2}):\;\underset{r_\mathcal{A}}{\text{max}} \;\eta_{\mathcal{E}}\{  r_\mathcal{A}\},
 \quad\text{subject to:}\; C1, C2, C7. \nonumber
\end{eqnarray}
Its solution is  provided via Lemma 2. 

 \begin{lemma}\label{lem:decodingRate}  
 {The global-optimal solution of (P2) is
   \begin{equation}\label{optrA}
r^\star_{\mathcal{A}} \triangleq \min\left(R_{\text{D}\mathcal{E}}, \frac{P_m-P_\text{fr}}{\rho_{_\text{d}}}\right).
\end{equation}
where 
$\rho_{_\text{d}}$ is measured in W/bps/Hz.}
 \end{lemma}
     
\begin{IEEEproof}
As observed from \eqref{AEE_e}, $ \eta_{\mathcal{E}}\{  r_\mathcal{A}\}$ is a decreasing function of $r_{\mathcal{A}}$ for $r_{\mathcal{A}} > R_{\mathcal{U}}$. On the other hand, 
as
$\frac{\partial \eta_{\mathcal{E}}\{  r_\mathcal{A}\}}{\partial r_\mathcal{A}} =  \frac{P_\text{fr} }{{({P_\text{fr}+ \rho_{_\text{d}}\hspace{.2mm} r_\mathcal{A} })}^2}>  0$, $\forall   r_{\mathcal{A}} \in   \left[0\,,  R_{\mathcal{A}}\right]\,$
we can observe that $ \eta_{\mathcal{E}}\{  r_\mathcal{A}\}$
is strictly increasing  function of $r_{\mathcal{A}}$ for $r_{\mathcal{A}} \leq R_{\mathcal{U}}$.  Therefore, the global-optimal solution of (P2) can be obtained by solving the boundary constraints on $r_{\mathcal{A}}$ as defined via $C1$, $C2$ and $C7$ for $\alpha=1$. On solving them and using \eqref{eq: C_LE}, we obtain \eqref{optrA}.
 \end{IEEEproof}
\subsection{Optimal Jamming Power}\label{sec:bh1} 
{For  
$\alpha=0$,  (P0) gets transformed to (P3), where $\eta_\mathcal{J}$ is to be maximised with $\mathcal{A}$ being in jamming mode.}
\begin{eqnarray}\label{P3}
 (\text{P3}):\;\underset{P_{\mathcal{J}}}{\text{max}} \quad\eta_{\mathcal{J}}\{P_{\mathcal{J}}\},
 \qquad \text{subject to:}\;  C3, C4, C7.\nonumber
\end{eqnarray}
Below, we present the key result leading to global-optimal solution of (P3).
\begin{lemma}\label{lem:jammingPower}   $\eta_{\mathcal{J}}\{P_{\mathcal{J}}\}$  is a psuedo-concave function of $P_{\mathcal{J}}$.
 \end{lemma}
\begin{IEEEproof}
  Using \eqref{eq: C_LJ} and \eqref{AEE_e}, it is found that $R_{\text{D}\mathcal{J}}$, the numerator of $\eta_{\mathcal{J}}\{P_{\mathcal{J}}\}$, is a concave
 function of $P_{\mathcal{J}}$ as 
$\frac{d^2R_{\text{D}\mathcal{J}}} {d P_{\mathcal{J}}^2}=-\dfrac{(\mathrm{g}_{_{\mathcal{S}\mathcal{U}}})^2\mathrm{g}_{_{\mathcal{S}\mathcal{A}}}P_{\mathcal{S}}\left(2\mathrm{g}_{_{\mathcal{A}\mathcal{U}}}P_{\mathcal{J}}+\mathrm{g}_{_{\mathcal{S}\mathcal{U}}}P_{\mathcal{S}}+2\sigma^{2}\right)}{\log\left(2\right)\left(\mathrm{g}_{_{\mathcal{A}\mathcal{U}}}P_{\mathcal{J}}+\sigma^{2}\right)^2\left(\mathrm{g}_{_{\mathcal{A}\mathcal{U}}}P_{\mathcal{J}}+\mathrm{g}_{_{\mathcal{S}\mathcal{U}}}P_{\mathcal{S}}+\sigma^{2}\right)^2} < 0,\forall \\ P_{\mathcal{J}}  \in  \left [0,   P{_{\mathcal{J}m}} \right]$. Next, since the denominator ${P_\text{ft}+(P_{\mathcal{J}}/\nu)}$ of $\eta_{\mathcal{J}}\{P_{\mathcal{J}}\}$ is an 
affine transformation of  
 $P_{\mathcal{J}}$ and the  ratio of concave function to positive affine function is a pseudo-concave function \cite[(Table 5.5)]{avriel2010generalized}, we notice that $\eta_{\mathcal{J}}\{P_{\mathcal{J}}\}$ is  pseudo-concave in $P_{\mathcal{J}}$. This completes the proof.
 \end{IEEEproof}

Lemma 3  implies that the global-optimal solution of (P3) as denoted by  $P_{\mathcal{J}}^\star$ is obtained by  solving $\frac{d\eta_{\mathcal{J}}\{P_{\mathcal{J}}\}} {d P_{\mathcal{J}}}=0$. Since, it is not possible to obtain the  explicit analytic solution  for $P_{\mathcal{J}}^\star$, we apply golden section (GS) search within lower bound $P^l=0$ and  upper bound  $ P^u =  \min(P_{\mathcal{J}m}, \nu(P_m-P_\text{ft}))$ as obtained using $C3$, $C4$  and $C7$ with $\alpha=0$. 
{Next,  we  analyse the complexity of  global-optimal solution of (P3) obtained numerically. In addition, we also introduce an approximate closed-form solution 
for $P_{\mathcal{J}}^\star$ to get further analytical insights.}
\subsubsection{Complexity Analysis}\label{sec:complexity}
\textcolor{red}{} 
 $P_{\mathcal{J}}^\star$ can be  found efficiently  in  few iterations $N $, which can be calculated noting  that the  GS search  terminates when $(P^u-P^l)(0.618)^N \leq  \epsilon$ where $\epsilon$ is the acceptable tolerance.
Thus, the order of complexity for the proposed solution, in Big O notation, is  $ O\left(\log\left(\frac{\min(P_{\mathcal{J}m}, \nu(P_m-P_\text{ft}))}{\epsilon}\right)\right)$.
 \subsubsection{Proposed Analytical Approximation}\label{sec:Approx}
 {Here, we propose an approximate closed-form solution for $P_{\mathcal{J}}^\star$ for the asymptotic case with
   $\gamma_{_{\mathcal{S}\mathcal{U}}} \gg 1$  and $\gamma_{_{\mathcal{A}\mathcal{U}}}\gg 1$ along with $ P_{\mathcal{J}}  \gg P_\text{ft}$ {\cite{EEE19}}.}   Under these approximations, using \eqref{eq: C_LJ} and \eqref{AEE_e}, objective function of (P3)  can be reduced to:
  \begin{eqnarray}\label{AEE_j_approx}
 \eta_{\mathcal{J}}\{P_{\mathcal{J}}\}\approx
\frac{\nu \log_2\left(\phi\right) }{P_{\mathcal{J}}}\; \text{with} \ \phi= \frac{P_{\mathcal{S}}\mathrm{g}_{_{\mathcal{S}\mathcal{U}}}P_{\mathcal{J}}\mathrm{g}_{_{\mathcal{A}\mathcal{U}}}}{\left(P_{\mathcal{S}}\mathrm{g}_{_{\mathcal{S}\mathcal{U}}}+P_{\mathcal{J}}\mathrm{g}_{_{\mathcal{A}\mathcal{U}}}\right)\sigma^2}.
  \end{eqnarray}
  Hereby, analytical approximation for  $P_{\mathcal{J}}^\star$ is obtained by
    first solving
           $ \frac{d\eta_{\mathcal{J}}\{P_{\mathcal{J}}\}} {d P_{\mathcal{J}}}\hspace{0mm}=0$, 
  which reduces to 
  $ \phi e^{\left(\frac{P_{\mathcal{J}}\mathrm{g}_{_{\mathcal{A}\mathcal{U}}}}{P_{\mathcal{S}}\mathrm{g}_{_{\mathcal{S}\mathcal{U}}}+P_{\mathcal{J}}\mathrm{g}_{_{\mathcal{A}\mathcal{U}}}}\right)}=\\e, 
 $ and then using the definition of Lambert function  \cite{LambertFunction} as:
  \begin{eqnarray}\label{P_j_approx}
 \widehat{P_{\mathcal{J}}^\star}\triangleq
 \frac{P_{\mathcal{S}}\mathrm{g}_{_{\mathcal{S}\mathcal{U}}}\mathrm{W}(\frac{e}{ \gamma_{_{\mathcal{S}\mathcal{U}}}})}{\mathrm{g}_{_{\mathcal{A}\mathcal{U}}}[1-\mathrm{W}(\frac{e}{ \gamma_{_{\mathcal{S}\mathcal{U}}}})]}.
 \end{eqnarray}
\subsection{Jointly Global-optimal Solution  $(\alpha^*, r^*_\mathcal{A}, P_{\mathcal{J}}^*)$ for (P0) } \label{sec:P0}

\begin{figure}[!t]
\centering
{{\includegraphics[width=3in]{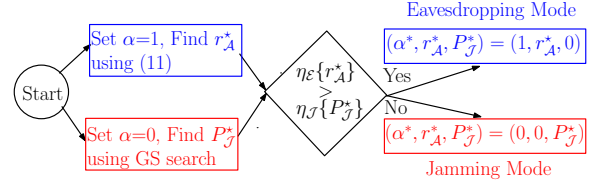} }}
 \centering\vspace{-2mm}
\caption{\small Graphical description for solution methodology.} 
    \label{fig: tree}
    \vspace{-1mm}
\end{figure}

{
As explained in Lemma 1 and Remark 1, optimal solutions of $r_\mathcal{A}$ and $P_{\mathcal{J}}$ are found using their decoupling for both optimal value of  $\alpha$ through (P2) and (P3) respectively.    
Lemma 2 and  3 further  prove the global-optimality of the solutions for $r_\mathcal{A}$ and $P_{\mathcal{J}}$.
Finally, we select the optimal setting of $\alpha$ out of 1 or 0 that gives the higher AEE between  $\eta_\mathcal{E}\{r^\star_\mathcal{A} \}$ and $\eta_\mathcal{J}\{P^\star_\mathcal{J} \}$.  This global-optimal value of $\alpha$ is obtained using \eqref{optalpha} while setting 
$r_\mathcal{A}=r^\star_\mathcal{A}$ and $P_{\mathcal{J}}=P_{\mathcal{J}}^\star$ for joint optimisation.}
Summarising the solution methodology, as also depicted by Fig. \ref{fig: tree},  the jointly global-optimal solution of (P0)  is:
\begin{equation}{ \label{global}}(\alpha^*,r^*_\mathcal{A}, P_{\mathcal{J}}^* ) = \begin{cases}( 1, r^\star_\mathcal{A}, 0) & 
\eta_\mathcal{E}\{r^\star_\mathcal{A} \}> \eta_\mathcal{J}\{P^\star_\mathcal{J} \}\\
( 0, 0, P_{\mathcal{J}}^\star) & \text{otherwise}. \\
\end{cases}\end{equation}
As $\alpha^*=1$ implies only eavesdropping, therefore, $P^*_{\mathcal{J}}=0 $. Similarly $r^*_\mathcal{A}= 0$ for $\alpha^*=0$ since   only jamming is performed.   
 \vspace{-2mm}
\section{Numerical Results} \label{result}
\begin{figure}[!t]
  \vspace{-1mm}\subfigure[Performance of $r^\star_\mathcal{A}$.]
    {{\includegraphics[width=1.95in]{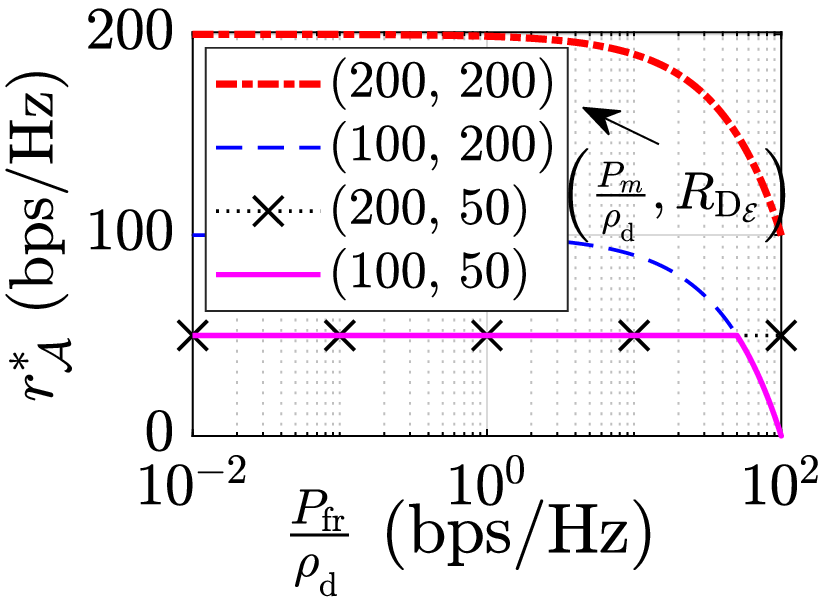} }}\hspace{-14mm}\vspace{-1mm} 
   \subfigure[ Nature of AEE $\eta_{\mathcal{J}}$ in $P_{\mathcal{J}}$.] 
   {{\includegraphics[width=1.95in]{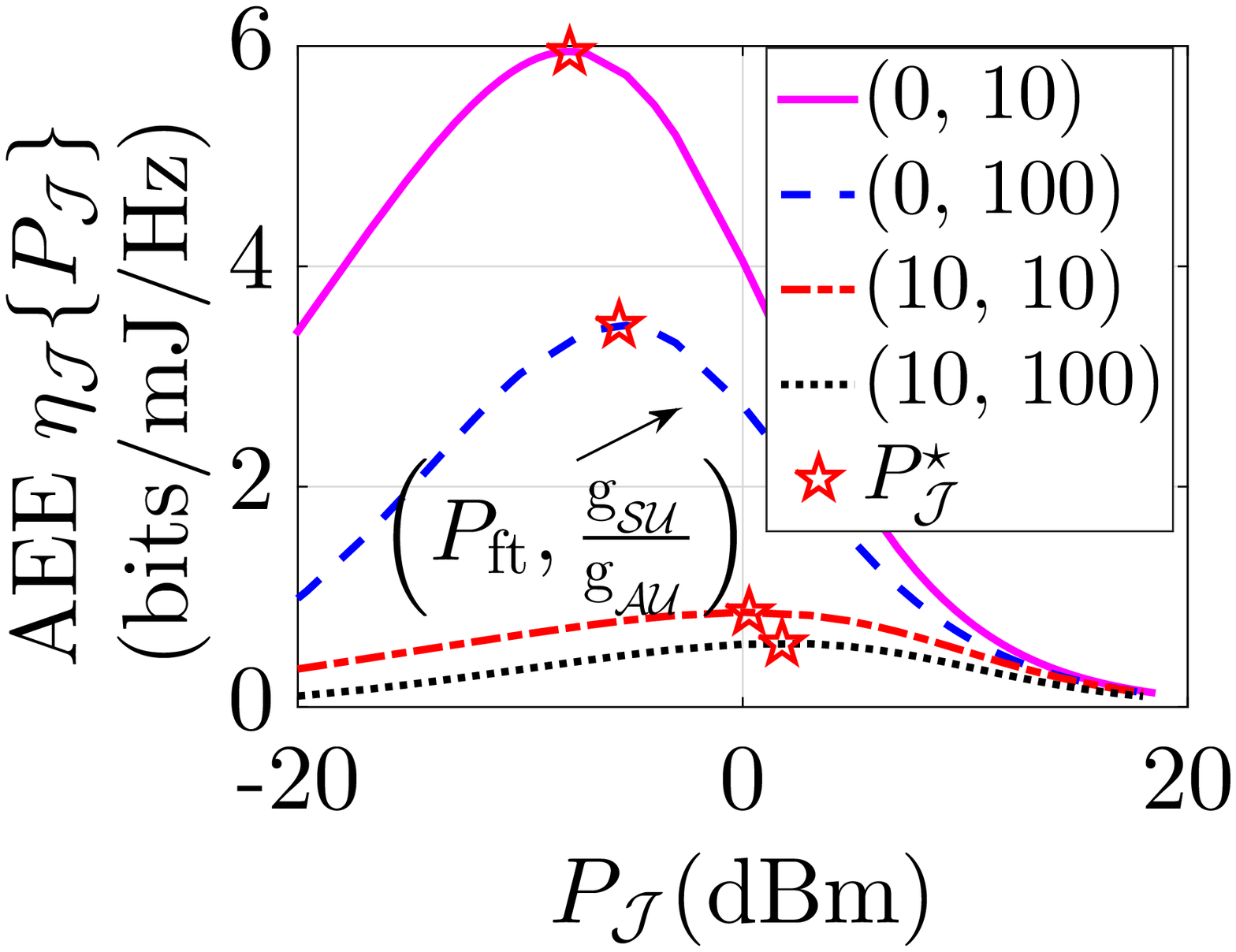} }}
   \hspace{-14mm}
       \caption{\small Analysis validation  and insights on optimal $r_\mathcal{A}$ and  $P_\mathcal{J}$. }%
    \label{fig:optimization} 
\end{figure}
{The default simulation parameters are: 
 $P_{\mathcal{S}}= 10$ dBm,  $P_{\mathcal{J}m}= 13$ dBm, $P_{{m}}= 13$ dBm,  $\mathrm{g}_{_{\mathcal{S}\mathcal{U}}}=-60$ dB,   $\mathrm{g}_{_{\mathcal{S}\mathcal{A}}}=\mathrm{g}_{_{\mathcal{A}\mathcal{U}}}=-70$ dB, $\sigma^2=-100$ dBm, $\nu=0.7$, $P_\text{ft}=P_\text{fr}=-0.33$ dBm, $\rho_{_\text{d}}=-10.33$ dBm/bps/Hz as the decoding power consumption per unit rate at $\mathcal{A}$  considering Mica2 mote
 and CC-1000 transceiver to represent low power IoT system
 \cite{Dilemma_Deepak,Access} and references therein.}
 
 First, through Fig. \ref{fig:optimization}(a),
 we explore the optimal eavesdropping performance by investigating the nature of   $r^\star_\mathcal{A}$   for different values of  $R_{\text{D}\mathcal{E}}$ and $\frac{P_m}{\rho_{_\text{d}}}$ in bps/Hz 
 by varying  
 $\frac{P_\text{fr}}{\rho_{_\text{d}}}$.
 The impact of different parameters can be categorized into two cases: (i) $R_{\text{D}\mathcal{E}}\geq  \frac{P_m}{\rho_{_\text{d}}} $, (ii) $R_{\text{D}\mathcal{E}} < \frac{P_m}{\rho_{_\text{d}}}$. For case (i), $r^\star_\mathcal{A}$ follows increasing trend with ${P_m} $ and decreasing  with ${P_\text{fr}}$ for a given ${\rho_{_\text{d}}}$.  A significant drop is  observed in $r^\star_\mathcal{A}$  beyond $\frac{P_\text{fr}}{\rho_{_\text{d}}}=10$ for the typical values, because  higher values of  $P_\text{fr}$ relative to $\rho_{_\text{d}}$ help to reduce $r^\star_\mathcal{A}$  significantly. 
 For case (ii),
 $r^\star_\mathcal{A}$ becomes independent of ${P_m} $ and depends on ${P_\text{fr}}$ for a given ${\rho_{_\text{d}}}$ only for $\frac{P_\text{fr}}{\rho_{_\text{d}}}>\frac{P_m}{\rho_{_\text{d}}}-R_{\text{D}\mathcal{E}}$. Otherwise $r^\star_\mathcal{A}$ remains equal to $R_{\text{D}\mathcal{E}}$ due to boundary constraints. For instance, with 
  $\frac{P_m}{\rho_{_\text{d}}}=100$ and
$R_{\text{D}\mathcal{E}}=50$ in Fig. \ref{fig:optimization}(a), $r^\star_\mathcal{A}$ remains equal to  $R_{\text{D}\mathcal{E}}$ till $\frac{P_\text{fr}}{\rho_{_\text{d}}}\leq 50$. Later on, it decreases with ${P_\text{fr}}$ for given $\rho_{_\text{d}}$.

 Next, via Fig \ref{fig:optimization}(b), we investigate the optimal jamming  
and validate the pseudo-concavity  of AEE $\eta_{\mathcal{J}}\{P_{\mathcal{J}}\}$   in  $P_{\mathcal{J}}$  for different values of $P_\text{ft}$ in dBm and channel gains ratio. 
 Here, $\mathrm{g}_{_{\mathcal{S}\mathcal{U}}}$  is kept fixed  and only $\mathrm{g}_{_{\mathcal{A}\mathcal{U}}}$  is varied. We note that $\eta_{\mathcal{J}}\{P_{\mathcal{J}}\}$ decreases with increased $P_\text{ft}$ because the latter leads to higher power consumption. Also, the underlying $P^\star_{\mathcal{J}}$ moves closer to  $P_{\mathcal{J}m}$ for further enhancement of secrecy degradation owing to its reduced  impact on  power consumption. Similarly, $\eta_{\mathcal{J}}\{P_{\mathcal{J}}\}$ decreases with decreased $\mathrm{g}_{_{\mathcal{A}\mathcal{U}}}$ due to reduced degraded secrecy rate and $P^\star_{\mathcal{J}}$ increases to strengthen the $\mathcal{A}$-$\mathcal{U}$ link.  Results  also show  that static power consumption has a significant impact in addition to  channel parameters on $\eta_{\mathcal{J}}\{P^\star_{\mathcal{J}}\}$  and hence, the former cannot be ignored unlike existing literature did.
 
\begin{figure}[!t]
\vspace{-3mm}
\centering
{{\includegraphics[width=3.1in]{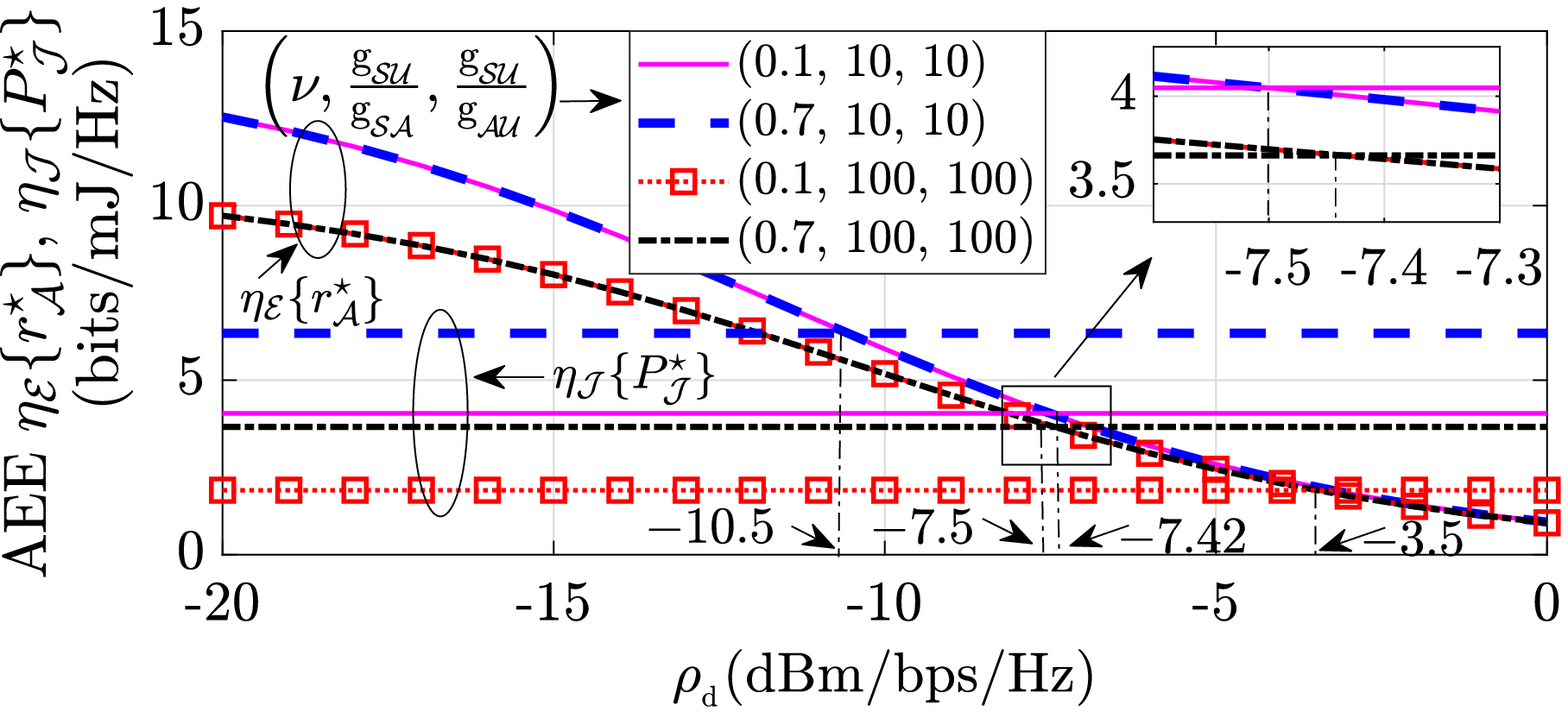} }}\vspace{-4mm}
\caption{\small Optimal mode switching based on $\eta_{\mathcal{E}}\{  r^\star_\mathcal{A}\}$ and $\eta_{\mathcal{J}}\{P^\star_{\mathcal{J}}\}$.}
    \label{fig:ModeSwitching}\vspace{-3mm}
\end{figure}

     Now we would like to provide novel  insights on the \textit{optimal mode switching} at $\mathcal{A}$ depending on   power consumption  and channel  parameters. We
    can note from \eqref{global} that $\eta_{\mathcal{E}}\{  r^\star_\mathcal{A}\}$
    and $\eta_{\mathcal{J}}\{P^\star_\mathcal{J}\}$ are deciding factors for optimal mode switching, and joint optimal AEE can be calculated as: 
    $\eta(\alpha^*,r^*_\mathcal{A}, P_{\mathcal{J}}^*)= \max(\eta_{\mathcal{E}}\{  r^\star_\mathcal{A}\},\eta_{\mathcal{J}}\{P^\star_\mathcal{J}\})$.  
    Thus, $\eta_{\mathcal{E}}\{r^\star_\mathcal{A}\}$ and $\eta_{\mathcal{J}}\{P^\star_{\mathcal{J}}\}$ are plotted in Fig. \ref{fig:ModeSwitching} for different values of channel gain ratio, $\nu$ and $\rho_{_\text{d}}$.
We observe that there exists a  switching threshold value of $\rho_{_\text{d}}$, below which $\eta_{\mathcal{E}}\{r^\star_\mathcal{A}\}$ is greater than $\eta_{\mathcal{J}}\{P^\star_{\mathcal{J}}\}$ resulting in  eavesdropping mode as optimal one. Otherwise, for $P_d$ values higher than this threshold, jamming is selected as the optimal mode. Results show that for a lower efficiency jammer with $\nu=0.1$, the channel conditions $
\{\frac{\mathrm{g}_{_{\mathcal{S}\mathcal{U}}}}{\mathrm{g}_{_{\mathcal{S}\mathcal{A}}}}=  \frac{\mathrm{g}_{_{\mathcal{S}\mathcal{U}}}}{\mathrm{g}_{_{\mathcal{A}\mathcal{U}}}}\}=\{10, 100\}$ lead to   eavesdropping  being the optimal mode for  $\rho_{_\text{d}}< \{-7.5, -3.5\}$dBm, respectively. However, for an efficient jammer with $\nu=0.7$, this switching threshold value decreases to $\{-10.5, -7.42\}$dBm for $\{\frac{\mathrm{g}_{_{\mathcal{S}\mathcal{U}}}}{\mathrm{g}_{_{\mathcal{S}\mathcal{A}}}}=  \frac{\mathrm{g}_{_{\mathcal{S}\mathcal{U}}}}{\mathrm{g}_{_{\mathcal{A}\mathcal{U}}}}\}=\{10, 100\}$, respectively. Thus, lower $\mathcal{A}$'s channel gains  $\left(\mathrm{g}_{_{\mathcal{S}\mathcal{A}}},\mathrm{g}_{_{\mathcal{A}\mathcal{U}}}\right)$ favor eavesdropping, whereas higher  $\rho_{_\text{d}}$ and  $\nu$ values favor jamming.
\vspace{-1mm}
\begin{figure}[!t]
\centering
{{\includegraphics[width=3.1in]{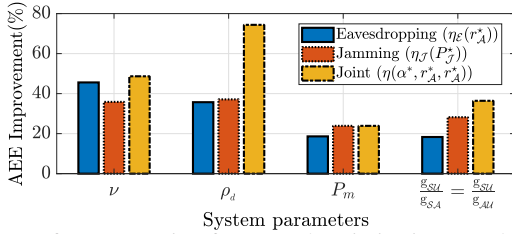} }}\vspace{-4mm}
\caption{\small Performance gain of proposed optimization over benchmark.} 
    \label{fig:Fig4}\vspace{-5mm}
\end{figure}

Finally, in Fig. \ref{fig:Fig4}, we compare the average AEE performance of three optimization schemes i) optimal  eavesdropping (P2)  ii) optimal jamming (P3), and  iii) joint optimization (P0) against a fixed benchmark scheme with $\alpha=0.5$, $P_{\mathcal{J}}=0$ dBm and $r_\mathcal{A}= \min\left(\log_{2}\left(1+ {\gamma_{_{\mathcal{S}\mathcal{A}}}}\right),\frac{P_m-P_\text{fr}}{\rho_{_\text{d}}}\right)$   which is  maximum possible eavesdropping rate.
 The average percentage ($\%$) improvement 
is calculated by first analyzing $\%$ gain in optimal AEE for different values of each parameter ($\nu$, $\rho_{_\text{d}}$, $P_m$, $\frac{\mathrm{g}_{_{\mathcal{S}\mathcal{U}}}}{\mathrm{g}_{_{\mathcal{S}\mathcal{A}}}}$, $\frac{\mathrm{g}_{_{\mathcal{S}\mathcal{U}}}}{\mathrm{g}_{_{\mathcal{A}\mathcal{U}}}}$) in its acceptable range and then taking  mean over these  respective gain values.  The corresponding average $\%$ improvement in AEE are plotted for different parameters    by varying $\nu$ from $10\%$ to $90\%$, $\rho_{_\text{d}}$ from $-20$ dBm/bit to $0$ dBm/bit, $P_m$  from $0$ dBm to $13$ dBm,  $\frac{\mathrm{g}_{_{\mathcal{S}\mathcal{U}}}}{\mathrm{g}_{_{\mathcal{S}\mathcal{A}}}}$ and $\frac{\mathrm{g}_{_{\mathcal{S}\mathcal{U}}}}{\mathrm{g}_{_{\mathcal{A}\mathcal{U}}}}$  from $1$ to $1000$. The average $\%$ improvement  as provided by optimal eavesdropping, jamming and joint schemes are $29.5\%$, $31.5\%$, $45\%$ respectively. It is to be noted that with $P_{\mathcal{J}}>0$ dBm in fixed benchmark scheme, the average percentage gain in AEE provided by proposed optimization schemes would have been even more higher.
 Thus, the joint optimization  provides a significant  improvement in AEE over benchmark scheme, and in general, the \textit{optimal jamming} is a better semi-adaptive scheme as compared to optimal eavesdropping.
\section{Concluding Remarks} \label{sec:Con}  We have derived the  jointly global-optimal  attacking mode selection and resource allocation for  maximizing the AEE.  Numerical investigations validate the key  metric characterizing  the non-trivial mode switching based  solution methodology (c.f. Fig. \ref{fig: tree})  for solving the otherwise non-convex problem.  Significant improvement of $45\%$ in average AEE is achieved using proposed joint optimization over the  benchmark.
  

\bibliographystyle{IEEEtran}


\end{document}